\documentclass[preprint, 12pt]{elsarticle}
\usepackage{graphicx}
\usepackage{dcolumn}
\usepackage{amssymb}
\usepackage{bm}
\usepackage{enumerate}
\journal{Astroparticle Physics}
\begin{document}
\begin{frontmatter}
%--------------------------------------------------------------------
%\def\Universita{Universit\`a}
%--------------------------------------------------------------------

%-------------------------------------------------------------------
%\date{, 19 October 2013; {\LaTeX-ed \today}}}
%--------------------------------------------------------------------

\title{Limiting Superluminal Electron and Neutrino Velocities using the 2010 Crab Nebula Flare and the IceCube PeV Neutrino Events}

\author{Floyd W. Stecker}
\address{Astrophysics Science Division\\NASA Goddard Space Flight Center\\Greenbelt, MD 20771, USA } 
\ead{floyd.w.stecker@nasa.gov}

\vspace{-1.5cm}
\begin{abstract}

The observation of two PeV-scale neutrino events reported by Ice Cube 
allows one to place constraints on Lorentz invariance violation (LIV) in the
neutrino sector. After first arguing that at least one of the PeV IceCube events was of extragalactic origin, I derive an upper limit for {\it the difference} between putative superluminal neutrino and electron velocities of $\le \sim 5.6 \times 10^{-19}$ in units where $c = 1$, confirming that the observed PeV neutrinos could have reached Earth from extragalactic sources. I further derive a new constraint on the superluminal electron velocity, obtained from the observation of synchrotron radiation from the Crab Nebula flare of September, 2010. The inference that the $>$ 1 GeV $\gamma$-rays from synchrotron emission in the flare were produced by electrons of energy up to $\sim 5.1$ PeV indicates the non-occurrence of vacuum \'{C}erenkov radiation by these electrons. This implies a new, strong constraint on superluminal electron velocities $\delta_e \le \sim 5 \times 10^{-21}$. It immediately follows that one then obtains an upper limit on the superluminal neutrino velocity {\it alone} of $\delta_{\nu} \le \sim 5.6 \times 10^{-19}$, many orders of magnitude better than the time-of-flight constraint from the SN1987A neutrino burst. However, if the electrons are {\it subluminal} the constraint on $|\delta_e| \le \sim 8 \times 10^{-17}$, obtained from the Crab Nebula $\gamma$-ray spectrum, places a
weaker constraint on superluminal neutrino velocity of $\delta_{\nu} \le \sim 8 \times 10^{-17}$.
\end{abstract}
\end{frontmatter}

\newpage

\section{Introduction}

Tests of Lorentz invariance violation (LIV) probe physics beyond the standard model, more specifically to probe the structure of space-time on the Planck scale. The Planck energy is the natural scale where it is expected that gravity may unify with the other three fundamental forces. While it is not possible to directly investigate space-time physics at the Planck energy of $\sim 10^{19}\,$GeV, many lower energy effects have been predicted to arise from the violation of Lorentz invariance. The subject of investigating LIV has therefore generated much interest in the particle physics community.

Various astrophysical observations using X-ray, $\gamma$-ray and cosmic ray data have been used to place limits on Lorentz violation (see e.g., Ref.~\cite{lb13}). Diffuse fluxes of high energy neutrinos produced both in our galaxy and in intergalactic space have long been predicted, but it was noted that at energies below several hundred GeV their fluxes would be swamped by the neutrinos produced by cosmic ray interactions in the atmosphere~\cite{st79}. Until recently there were no reported detections of cosmic-ray neutrinos. However, the IceCube collaboration has now reported the first observation of two PeV energy neutrinos, giving a signal $\sim 3\sigma$ above the
atmospheric background~\cite{aa13a}. It is most likely that these neutrinos
are evidence of a new neutrino flux component above that expected from
atmospheric cosmic ray secondaries and that such a component would be of 
extraterrestrial origin~\cite{aa13a,la13}. 

While more data are desirable, there are four indications that all or most of these neutrinos are extragalactic in origin: (1) The IceCube collaboration has reported 18 more events produced by neutrinos with energies above 0.1 PeV, $\sim$ 4 $\sigma$ above the expected atmospheric background. The distribution in arrival direction of all 20 events is consistent with isotropy; there is no marked enhancement in the galactic plane~\cite{ha13,aa13b}, although it has been argued that a subset of these events might be of galactic origin~\cite{ra13a,ah13}. (2) The implied peak in the energy spectrum of these neutrinos may be indicative of photopion production followed by pion decay ~\cite{wi13,ch13,ki13} such as expected in AGN cores~\cite{st91,st13a} and GRBs~\cite{li13,ra13b}. (3) The diffuse galactic neutrino flux~\cite{st79} is expected to be well below that implied by the implied Ice Cube flux. (4) At least one of the $\sim$ PeV neutrinos came from a direction off of the galactic plane. Even the existence of this one extragalactic neutrino event is
enough to place new constraints on LIV.

Limits on superluminal neutrino velocity $\delta_{\nu} = v_{\nu}-1 \le \sim 10^{-5}$ have been obtained directly from terrestrial time-of-flight measurements~\cite{ad07,dr13}. In addition, time-of-flight constraints from the detection of a multi-MeV neutrino burst from supernova 1987A~\cite{bi87,hi87} yielded the constraint  $\delta_{\nu} \le 2 \times 10^{-9}$~\cite{lo87}.
A comparison of atmospheric neutrino spectra with theoretical spectra expected from the change in the pion decay rate if neutrinos are superluminal has yielded the indirect constraint $\delta_{\nu} \le \cal{O}$(10$^{-13}$)~\cite{co12}. New IceCube observations, together with new constraints on superluminal electron velocities derived from $\gamma$-ray observations of the September, 2010 Crab Nebula flare using the Large Area Telescope on the Fermi Gamma Ray Space Telescope, now allow one to place stronger constraints on LIV in both the electron and the neutrino sectors. 

\section{Neutrino Energy Loss}

Calliday and Kosteleck\'{y}~\cite{ck98} proposed an effective field theory framework for quantifying and cataloging the empirical effects of small violations of {\cal CPT} and Lorentz invariance known
as the standard model extension (SME). The SME is based on the introduction of small Lorentz and {\cal CPT} violating perturbations in the individual free particle Lagranians. Coleman and Glashow~\cite{co99} have presented a simplified formalism, assuming rotational invariance, wherein particle interactions that violate Lorentz invariance can be modified in terms of the maximum attainable velocities (MAVs) of the various particles involved. Thus superluminal particle velocities can be directly related to Lorentz invariance violation. Cohen and Glashow~\cite{cg11} point out that if $\delta_{\nu} > 0$, three energy loss processes that are otherwise kinematically forbidden, would be allowed even {\it in vacuo}, {\it viz.} (a) vacuum neutrino \'{C}erenkov radiation $(\nu \to \nu \, \gamma)$, (b) "neutrino splitting" $(\nu \to \nu\, \nu \,\bar\nu)$ and (c) vacuum electron-positron emission $(\nu \to \nu \,e^+\, e^-)$ (see also~\cite{li13}). Of these processes, electron-positron pair emission is the most dominant, leading to the fastest energy loss. The MAVs of
the various neutrino flavors are found to be within a factor of less than $10^{-20}$ of each other, given the results from neutrino oscillation experiments~\cite{co99,cg11}.

We now define $\delta_{\nu} = v_{\nu} - 1$, $\delta_{e} = v_{e} - 1$, where c = 1 is the low energy velocity of light {\it in vacuo} and the $v$'s are the MAVs of the $\nu$'s and electrons. (N.B.: The definition of $\delta$ used here is {\it half} that used in Refs.~\cite{co99} and~\cite{cg11}, but is consistent with that used in Ref.~\cite{sg01}.) For $\delta_{\nu} \ge \delta_e \ge 0$ and defining $\delta_{\nu e} \equiv \delta_{\nu} - \delta_e$, the process $\nu \to \nu\, e^+\, e^-$ is kinematically allowed  provided that~\cite{co99,sg01} 
\begin{equation}
E_{\nu} \ge m_e\sqrt{2/\delta_{\nu e}}
\end{equation} 
 
Using the results of~\cite{cg11}, the attenuation length for this process is determined by the differential equation for energy loss by pair production (taking $\hbar = c = 1$)
\begin{equation}
\frac{dE}{dx} =\frac{25}{448}\frac{G_F^2 E^6 (2\delta_{\nu e})^3}{192\,\pi^3}, 
\end{equation} 
which leads to neutrinos having a terminal energy $E_T$ after traveling a distance $L$ given by 
~\cite{cg11}
\begin{equation}
E_T^{-5} = \frac{125}{448}\frac{G_F^2 (2\delta_{\nu e})^3}{192\,\pi^3} = 9.2 \times 10^{-15}\delta_{\nu e}^3 {\rm GeV^4} L({\rm f)}.
\label{terminal}
\end{equation}
\\
A similar result has been obtained in Ref.~\cite{be12} and has been generalized in Ref.~\cite{ca12}.
We note that 1 kiloparsec (kpc) = 3.085$\times 10^{34}$ f.
It then follows that for a terminal energy of $10^6$ GeV, a superluminal neutrino with $\sim$1 PeV
energy can have traveled a distance of
\begin{equation}
L({\rm kpc}) \le 9.4 \times 10^{-52} \delta_{\nu e}^{-3}
\label{pathlength}
\end{equation}

Then, taking $\delta_{\nu e} = 5.6 \times 10^{-19}$ from equation (1) for $E_{\nu}$ = 1 PeV, it follows from equation (4) that superluminal neutrinos with multi-PeV or greater energies that survive to a terminal energy $E_T \sim 1$ PeV cannot have propagated over a distance $\ge$ $\sim 32$ Mpc in the presence of energy loss by pair emission. This distance is of order of the size of the local supercluster of galaxies\footnote{Recently a new IceCube neutrino event named "big bird" was detected with an energy of 2.1 PeV with a 15\% error~\cite{kl13}. The arrival direction of this event was not released.}
\footnote{Very recently, it has been shown that with fluctuations and changes in the 
$e^+e^-$ radiation rate during propagation taken into account, some neutrinos with energy $\sim$ 1 PeV can have survived over longer path lengths than that given by equation (\ref{pathlength}). The cutoff energy in the resulting neutrino spectrum can thus be larger than the terminal 
energy given by equation (\ref{terminal})~\cite{ma14}. Assuming a cutoff energy of 
$1 PeV \sim 2E_T$~\cite{ma14,ma13}, the resulting pathlength obtained would be $2^{5}$ times 
larger than that given by equation (\ref{pathlength}) {\it viz.} $\sim$ 1 Gpc.}

The neutrino event that originated clearly away from the galactic plane had a measured cascade energy in IceCube of 1.14 PeV. The uncertainty in this determination is 15\% ~\cite{aa13c}.  We can therefore use 1 PeV as a conservative value for the its energy. In Section 1 we have given reasons as to why this $\sim$PeV neutrino was of extragalactic origin. Since the vast majority of candidate extragalactic candidate sources lie beyond the local supercluster, it is thus probable that the value of $\sim 5.6 \times 10^{-19}$ is a a valid upper limit on $\delta_{\nu e}$, and most likely a conservative one. 

The value of $\sim 1$ PeV for the neutrino events assumes that all of the energy of the incoming
neutrino is deposited in the Ice Cube detector. This is the case for charged current (CC) interactions. An IceCube cascade event may also be produced by a neutral current (NC) interaction. In that case, owing to the small average inelasticity of the NC interaction, $<y>\ \simeq 0.26$~\cite{ga96} producing the observed cascade, the initial neutrino energy, $E_{\nu}$ can be significantly greater than the energy deposited in the cascade. It can, in fact, be several PeV. Thus, in the NC case, equation (1) would yield a smaller value for the upper limit on $\delta_{\nu e}$. However, the probability for NC events is smaller than for CC events because the NC cross section is smaller and also because of the dependence of the expected event rate on the initial neutrino energy spectrum. 

\section{Limits on Superluminal Electron and Neutrino Velocities}

It is important to note that what we obtained in the previous section is a limit on {\it the difference} between the neutrino and electron velocities, $\delta_{\nu e}$, not on the neutrino velocity itself. We derived a conservative upper limit on $\delta_{\nu e}$. However, our final goal is to derive the more physically fundamental upper limits on the superluminal electron and neutrino velocities {\it separately}, {\it i.e.}, $\delta_{\nu}$ and $\delta_{e}$~\cite{note}.

\subsection{The Superluminal Electron Velocity Constraint}

Previous indirect constraints on $\delta_e \simeq \cal{O}$(10$^{-15}$)\cite{al06}. Here we use $\gamma$-ray spectral data from the September, 2010 flare of the Crab Nebula showing the acceleration of electrons in the flare to multi-PeV energies and allowing the deduction of best LIV constraints in the electron sector to date.

Synchrotron $\gamma$-rays from this strong flare were observed by the Large Area Telescope on the Fermi Gamma Ray Space Telescope up to an energy $>$ 1 GeV~\cite{ab11}.This has provided evidence for the acceleration of electrons in the nebula up to PeV energies. Synchrotron emission occurs when relativistic electrons are accelerated in magnetic fields~\cite{sc49}. The characteristic $\gamma$-ray energy produced by electrons of energy $E_e$ in a magnetic field of strength $B_{\perp}$ perpendicular to the motion of the electron is given by ({\it e.g.}~\cite{st71})
\begin{equation}
E_{\gamma} = 1.9 \times 10^{-11} B_{\perp} E_e^2
\end{equation}
\noindent where the energies are in GeV and the magnetic field strength is in gauss. Conservatively
taking a maximum magnetic field strength for the inner nebula of $2 \times 10^{-3}$ G~\cite{he95}, we find a characteristic electron energy $E_e >$ 5.1 PeV. The implication that  electrons of this energy have not been eliminated by the emission of vacuum \'{C}erenkov radiation places an upper limit on $\delta_e$ given by~\cite{co99,sg01}
\begin{equation}
\delta_e \le \frac{1}{2(E_{e}/m_e)^2} \simeq 5 \times 10^{-21}
\label{delta-e}.
\end{equation}
This new constraint is five orders of magnitude stronger than the direct constraint given in Ref.~\cite{sg01} (see also \cite{note2}).  

\subsection{Superluminal Neutrino Velocity Constraint Assuming $\delta_e \ge 0$}

Since eq. (\ref{delta-e}) implies that $\delta_{\nu e} \gg \delta_e$, we find that $\delta_{\nu} \simeq \delta_{\nu e} \sim 5.6 \times 10^{-19}$, almost ten orders of magnitude better than the time-of-flight constraint from the SN1987A neutrino burst~\cite{lo87} and more than five orders of magnitude better than the constraint obtained from the study of atmospheric neutrino spectra~\cite{co12}. Our new constraints apply directly to the dimension-4 operators in the SME; $c^e_{TT} \equiv -\delta_e$ and $\mathaccent'27 c^{(4)} \equiv -\delta_\nu$ (see tables D6 and D19 of Ref.~\cite{kr13}). Such constraints have important implications for quantum gravity models and Planck scale physics.

\subsection{Superluminal Neutrino Velocity Constraint Assuming $\delta_e \le 0$}

If, however, we allow the possibility that the electron velocities may be {\it subluminal} we get a looser, but still significant, constraint on superluminal neutrino velocities. In this case, the
{\it in vacuo} decay of photons into electron-positron pairs is kinematically allowed for photons
with energies exceeding a maximum energy given by \cite{sg01}

\begin{equation}
E_{\gamma, \rm max}= m_e\,\sqrt{2/|\delta_{e}|}\;. 
\end{equation}

The  decay would take place rapidly, so that photons with energies exceeding 
$E_{\gamma, \rm max}$ could not be observed either in the laboratory or as cosmic 
rays. From the fact that photons have been observed with energies   
$E_{\gamma} \ge$ 80~TeV from the Crab nebula \cite{ah04}, we deduce for this 
case that $E_{\rm max}\ge 80\;$TeV, or that the magnitude of the negative value for 
$|\delta_e|$ is less than $8\times 10^{-17}$. This value is comparable to the
constraint given in \cite{note2}.

In this case then, with $\delta_{\nu e}$ less than $\sim 5.6 \times 10^{-19}$ as determined
from eq. (1), we find 

\begin{equation}
\delta_{\nu} = \delta_{\nu e} + |\delta_e| \simeq |\delta_e| = 8\times 10^{-17}.
\end{equation}

\section*{Acknowledgment} I thank Francis Halzen for discussions regarding the IceCube results. I also thank Alice Harding and Alan Kosteleck\'{y} for helpful discussions.


\begin{thebibliography}{99}

\bibitem{lb13}  S.~Liberati, Class. Quantum Grav. {\bf 30}, 133001 (2013).

\bibitem{st79} F. W. Stecker, Astrophys. J. {\bf 228}, 919 (1979).

\bibitem{aa13a} M. G. Aartsen et al., Phys. Rev. Lett. {\bf 111}, 021103.

\bibitem{la13} R. Laha, et al.,  Phys. Rev. D {\bf 88}, 043009 (2013).

\bibitem{ha13} F. Halzen, {\it Proc. 33rd Intl. Cosmic Ray Conf., Rio de Janeiro}, 2013.

\bibitem{aa13b} M. G. Aartsen et al., Science {\bf 342}, 242856 (2013).

\bibitem{ra13a} S. Razzaque, Phys. Rev. D {\bf 88}, 081302.

\bibitem{ah13} M. Ahlers and K. Murase, e-print arXiv:1309.4077.

\bibitem{wi13} W. Winter,  Phys. Rev. D {\bf 88}, 083007 (2013).

\bibitem{ch13} Cholis and D. Hooper, JCAP {\bf 1306} 030 (2013).

\bibitem{ki13} M. D. Kistler, T. Stanev and H. Yuksel, e-print arXiv:1301.1703.

\bibitem{st91} F. W. Stecker, C. Done, M.H. Salamon and P. Sommers,
Phys. Rev. Lett. {\bf 66}, 2697 (1991).

\bibitem{st13a} F. W. Stecker, Phys. Rev. D {\bf 88}, 047301 (2013).

\bibitem{li13} R.-Y. Liu and X.-Y. Wang, Astrophys. J. {\bf 766}, 73 (2013).

\bibitem{ra13b} S. Razzaque, Phys. Rev. D {\bf 88}, 103003 (2013).

\bibitem{ad07} P. Adamson et al. (MINOS), Phys. Rev. D {\bf 76}, 072005. 
(2007). 

\bibitem{dr13} M. Dracos (OPERA), Nucl. Phys. B (Proc. Suppl.) {\bf 235}, 283 (2013).
 
\bibitem{bi87} R. Bionta, G. Blewitt, C. Bratton, D. Casper, and A. Ciocio et al., Phys. Rev. Lett. {\bf 58}, 1494 (1987). 

\bibitem{hi87} K. Hirata et al. (KAMIOKANDE-II), Phys. Rev. Lett. {\bf 58}, 1490 (1987). 

\bibitem{lo87} M. J. Longo, Phys. Rev. D {\bf 36}, 3276 (1987).

\bibitem{co12} R. Cowsik, T. Madziwa-Nussinov, S. Nussinov and U. Sarkar, Phys. Rev. D {\bf 86}, 045024 (2012).

\bibitem{ck98} D. Calladay and V. A. Kosteleck\'{y}, Phys. Rev. D {\bf 58}, 116002 (1998).

\bibitem{co99} S. R. Coleman and S. L. Glashow, Phys. Rev. D {\bf 59}, 
116008 (1999).

\bibitem{cg11} A.~G.~Cohen and S.~L.~Glashow, Phys.\ Rev.\ Lett.\  {\bf 107}, 181803 (2011).

\bibitem{sg01} F. W. Stecker and S. L. Glashow, Astropart. Phys. {\bf 16}, 97 (2001).

\bibitem{be12} F. Bezukov and H. M. Lee, Phys. Rev. D {\bf 85}, 031901 (2012).

\bibitem{ca12} J. M. Carmona, J. L. Cort\'{e}s, and D. Maz\'{o}n, Phys. Rev. D {\bf 85}, 113001 (2012).

\bibitem{aa13c} M. G. Aartsen et al., e-print arXiv:1311.4767.

\bibitem{kl13} S. Klein, in Proc. 33rd Intl. Cosmic Ray Conf., e-print arXiv:1311.6519.

\bibitem{ma14} Maz\'{o}n, e-print arXiv:1401.2964.

\bibitem{ma13} L. Maccione, S. Liberati, and D. Mattingly, JCAP, 1303, 039 (2013).

\bibitem{ga96} R. Gandhi, C. Quigg, M. H. Reno and I. Sarcevic, Astropart. Phys. {\bf 5}, 81 (1996).

\bibitem{note} Recently Borriello et al., Phys. Rev. D {\bf 87}, 116009 (2013), arrived at a similar constraint on $\delta_{\nu}$ based on the critical {\it assumption} that $\delta_{\nu} \gg \delta_e$. That assumption is proven in this paper.

\bibitem{al06} B.~Altschul, Phys.\ Rev.\ D {\bf 74}, 083003 (2006).

\bibitem{ab11} A. A. Abdo, et al., Science {\bf 331},739 (2010).

\bibitem{sc49} J. Schwinger, Phys. Rev. {\bf 75}, 1912 (1949).

\bibitem{st71} F. W. Stecker, Cosmic Gamma Rays, NASA-SP249 (U.S. Govt. Printing Office, Washington, DC, 1971).

\bibitem{he95} J. J. Hester et al., Astrophys. J. {\bf 448}, 240 (1995).

\bibitem{note2} A new limit based on radio spectroscopy of atomic dysprosium (M.A. Hohensee et al.,
Phys.\ Rev.\ Lett.\ {\bf 111} 050401 (2013)) has constrained LIV anisotropies in the electron sector to $\cal{O}$$(10^{-17}$).

\bibitem{kr13} V. A. Kosteleck\'{y} and N. Russell, e-print arXiv:0801.0287v6 (2013).

\bibitem{ah04} F. Aharonian et al., Astrophys, J. {\bf 614}, 897 (2004).


\end{thebibliography}
\end{document}